\documentstyle[pra,aps,twocolumn]{revtex}
\begin{document}
\draft
\title{Deterministic Quantum State Transformations}
\author{Anthony Chefles}
\address{Department of Physical Sciences, University of Hertfordshire \\
       Hatfield AL10 9AB, Herts, UK \\ email: A.Chefles@herts.ac.uk}
\input epsf
\epsfverbosetrue
\maketitle

\begin{abstract}
We derive a necessary condition for the existence of a
completely-positive, linear, trace-preserving map which
deterministically transforms one finite set of pure quantum states
into another.  This condition is also sufficient for
linearly-independent initial states.  We also examine the issue of
quantum coherence, that is, when such operations maintain the
purity of superpositions.  If, under any deterministic
transformation from one linearly-independent set to another, even
a single complete superposition maintains its purity, the initial
and final states are related by a unitary transformation.
 \\ *
\end{abstract}

Some of the most intriguing issues in quantum theory are,
ultimately, concerned with whether or not some given, desirable,
transformation of the state of a quantum mechanical system is
physically possible.  The general structure of
physically-realisable transformations is now well-established. Any
permissible transformation of the state ${\rho}$ of a quantum
system is represented by a completely-positive, linear,
trace-preserving (CPLTP) map ${\cal L}:{\rho}{\rightarrow}{\cal
L}({\rho})$\cite{Kraus,Schumacher}. Conversely, any such map
represents a transformation that is at least, in principle,
realisable.

The so-called first representation theorem\cite{Kraus} states that
the CPLTP conditions are equivalent to the requirement that ${\cal
L}$ can be represented as
\begin{equation}
{\cal L}({\rho})=\sum_{k}A_{k}{\rho}A^{\dagger}_{k},
\end{equation}
where the transformation operators $A_{k}$ may be any set of
linear operators which satisfy a resolution of the identity,
\begin{equation}
\sum_{k}A^{\dagger}_{k}A_{k}=1,
\end{equation}
and `1' is the identity operator.  This representation is also
known as the operator-sum representation\cite{Schumacher}. The
number of transformation operators $A_{k}$ is, in principle,
unrestricted, although if $D$ is the dimension of the Hilbert
space of the system, then any CPLTP map can be realised using no
more than $D^{2}$ transformation operators.

Quantum theory has both deterministic and probabilistic aspects. The archetypical example
of deterministic behaviour in quantum mechanics is the unitary evolution of a closed
quantum system according to the Schr\"odinger equation.  This evolution transforms every
pure quantum state into another pure state with certainty.  The final state can be
determined through knowledge of the initial conditions and the Hamiltonian, which gives
the law of evolution, just as in classical mechanics. CPLTP maps giving unitary evolution
correspond to only one of the transformation operators $A_{k}$ being non-zero.  Unitarity
is then implied by Eq. (2).

At perhaps the other extreme is the state transformation induced
by the measurement process.  According to von Neumann's
prescription\cite{Von}, in a maximal measurement the
transformation operators form a complete set of orthogonal, rank
one projectors. Each value of the index $k$ corresponds to a
distinguishable outcome of the measurement, these having
probability ${\mathrm Tr}({\rho}A^{\dagger}_{k}A_{k})$. When the
transformation operators are non-orthogonal, the operation is a
generalised measurement, whose outcome probabilities are still
given by this trace formula. Pure states are generally transformed
into mixed states during the measurement process, effectively
representing the uncertainty in the final quantum state.

In this Letter, we examine the conditions under which one set of
pure quantum states can be deterministically transformed into
another. A simple necessary, and for linearly-independent initial
states, sufficient condition is derived.  From this condition, it
is shown that each pair of final states must be less
distinguishable than their initial counterparts.

An important question is, when do such transformations preserve
the purity of superpositions?  We derive the following interesting
property of deterministic transformations between two sets of
linearly-independent states.  If such a transformation preserves
the purity of even a single, complete superposition of the initial
states, the initial and final states must be related by a unitary
transformation.

The issue of determinism in quantum state transformations can be
examined within the following scenario.  One party, Alice,
prepares a quantum system in one of the $N$ pure states
$|{\psi}_{j}^{1}{\rangle}$, where $j=1,{\ldots},N$, which span a
Hilbert space ${\cal H}$. She then passes the system onto her
colleague Bob. Bob does not know which state Alice prepared,
although he does know what the possible states
$|{\psi}_{j}^{1}{\rangle}$ are. His task is to devise a means of
transforming, with unit probability, each possible initial state
into a corresponding member of a set of final states
$|{\psi}_{j}^{2}{\rangle}$.  The question is, for fixed initial
and final states, under what conditions is such a transformation
physically possible?

From Eq. (1), it follows that
\begin{equation}
|{\psi}_{j}^{2}{\rangle}{\langle}{\psi}_{j}^{2}|=\sum_{k}A_{k}|{\psi}_{j}^{1}{\rangle}{\langle}{\psi}_{j}^{1}|A^{\dagger}_{k},
\end{equation}
or equivalently, that
\begin{equation}
A_{k}|{\psi}_{j}^{1}{\rangle}=c_{jk}|{\psi}_{j}^{2}{\rangle},
\end{equation}
for some complex coefficients $c_{jk}$.

We assume that the final states also lie in ${\cal H}$. This is
results in no loss of generality, since any set of $N$ pure states
can be brought into this space reversibly using a unitary
transformation.  This can be seen from the linearity of Eq. (4),
which implies that the space spanned by the final states cannot
have higher dimension than ${\cal H}$.  We may also, again without
loss of generality, take ${\cal H}$ to be the Hilbert space of the
entire system. Both of these assumptions are made for reasons of
notational convenience.

The linearity of the transformation operators $A_{k}$ implies
that, if the $|{\psi}_{j}^{1}{\rangle}$ are linearly-dependent,
then the problem is over-constrained.  It is easy to see why: if,
for example, $|{\psi}_{1}^{1}{\rangle}$ is a superposition of the
remaining states, that is,
$|{\psi}_{1}^{1}{\rangle}=\sum_{j=2}^{N}q_{j}|{\psi}_{j}^{1}{\rangle}$,
then how it is transformed is completely determined by how the
other states are transformed through Eq. (4).  Thus, if the
initial states span a Hilbert space ${\cal H}$, we can examine the
problem fully for a spanning, linearly-independent subset of the
states, and then derive consistency conditions which must be
satisfied for a larger, linearly-dependent set.  We shall examine
such issues later.

We wish to know when there exists a CPLTP map which
deterministically transforms one specified set of pure states into
another.  A necessary, and for linearly-independent initial
states, sufficient condition for the existence of such a map is
given by

\newtheorem{theorem}{Theorem}
\begin{theorem} Let $\{|{\psi}_{j}^{1}{\rangle}\}$ be a set of $N$ pure quantum states
spanning a D-dimensional complex Hilbert space ${\cal H}$.  Let
$\{|{\psi}_{j}^{2}{\rangle}\}$ be another set of $N$ pure states
lying in ${\cal H}$.  We also define a further $N$-dimensional
complex Hilbert space ${\cal H}_{N}$, for which we will write
operators and vectors in bold. Consider the Hermitian operator
${\mathbf M}=\{{\mu}_{jj'}\}$ on ${\cal H}_{N}$, with the matrix
representation
\begin{equation}
{\mu}_{jj'}=\frac{{\langle}{\psi}^{1}_{j'}|{\psi}^{1}_{j}{\rangle}}{{\langle}{\psi}^{2}_{j'}|{\psi}^{2}_{j}{\rangle}}.
\end{equation}
A necessary condition for the existence of a CPLTP map
transforming each $|{\psi}_{j}^{1}{\rangle}$ into
$|{\psi}_{j}^{2}{\rangle}$ is that ${\mathbf M}$ is positive. This
condition is also sufficient if the $|{\psi}_{j}^{1}{\rangle}$ are
linearly-independent.
\end{theorem}

{\noindent}{\bf Proof:} The necessity of this condition being
fulfilled follows readily from Eqs. (2) and (4).  From these
equations, we see that
\begin{equation}
{\mu}_{jj'}=\sum_{k}c_{j'k}^{*}c_{jk},
\end{equation}
from which it follows that, for any vector ${\mathbf v}=\{v_{j}\}$
in ${\cal H}_{N}$,
\begin{equation}
{\langle}{\mathbf v},{\mathbf M}{\mathbf
v}{\rangle}=\sum_{k}\left|\sum_{j}c_{jk}v_{j}\right|^{2}{\ge}0,
\end{equation}
proving that ${\mathbf M}$ must be positive.

To prove sufficiency in the case of linearly-independent initial
states, we make use of the fact that if ${\mathbf M}$ is positive,
then it may be written as
\begin{equation}
{\mathbf M}={\mathbf C}{\mathbf C^{\dagger}},
\end{equation}
for some other operator ${\mathbf C}=\{c_{jk}\}$ on ${\cal
H}_{N}$.

If, and only if, the $|{\psi}^{1}_{j}{\rangle}$ are
linearly-independent,  then we can define the reciprocal states
$|{\psi}^{1{\perp}}_{j}{\rangle}$.  The reciprocal state
$|{\psi}^{1{\perp}}_{j}{\rangle}$ is the state in ${\cal H}$ which
is orthogonal to all $|{\psi}^{1}_{j'}{\rangle}$ for $j{\neq}j'$.

Using these quantities, we can define the transformation operators
\begin{equation}
A_{k}=\sum_{j}\frac{c_{jk}}{{\langle}{\psi}^{1{\perp}}_{j}|{\psi}^{1}_{j}{\rangle}}|{\psi}^{2}_{j}{\rangle}{\langle}{\psi}^{1{\perp}}_{j}|.
\end{equation}
Acting upon the state $|{\psi}^{1}_{j}{\rangle}$ with this
operator gives the desired expression Eq. (4).  The condition that
the operators $A_{k}^{\dagger}A_{k}$ form the resolution of the
identity in Eq. (2) can also be seen by evaluating
$\sum_{k}A_{k}^{\dagger}A_{k}$ and making use of Eqs. (5) and (8),
giving
\begin{equation}
\sum_{k}A_{k}^{\dagger}A_{k}=\sum_{jj'}\frac{{\langle}{\psi}^{1}_{j'}|{\psi}^{1}_{j}{\rangle}}{{\langle}{\psi}^{1}_{j'}|{\psi}^{1{\perp}}_{j'}{\rangle}{\langle}{\psi}^{1{\perp}}_{j}|{\psi}^{1}_{j}{\rangle}}|{\psi}^{1{\perp}}_{j'}{\rangle}{\langle}{\psi}^{1{\perp}}_{j}|.
\end{equation}
That this operator is the identity may be seen from the fact that
we may write the identity operator as
\begin{equation}
1=\sum_{j}\frac{|{\psi}_{j}^{1}{\rangle}{\langle}{\psi}^{1{\perp}}_{j}|}{{\langle}{\psi}^{1{\perp}}_{j}|{\psi}_{j}^{1}{\rangle}}=\sum_{j}\frac{|{\psi}_{j}^{1{\perp}}{\rangle}{\langle}{\psi}^{1}_{j}|}{{\langle}{\psi}^{1}_{j}|{\psi}_{j}^{1{\perp}}{\rangle}}.
\end{equation}
The product of these expressions is equal to
$\sum_{k}A_{k}^{\dagger}A_{k}$, as can be verified using Eq. (10),
and is clearly also equal to the identity operator. This completes
the proof.

For linearly-independent states, $N=D$.  It follows then from Eq.
(8) that the rank of the operator ${\mathbf C}^{\dagger}$ need not
exceed $D$. This implies that any deterministic transformation of
a set of $D$ linearly-independent pure states requires no more
than $D$ transformation operators, rather than general minimum of
$D^{2}$.

Note that if the initial states are orthogonal, ${\mathbf M}$ is
simply the identity operator, which is obviously positive
irrespective of the final states.  Thus, as we would expect,
orthogonal states can be transformed into any other states.

One important consequence of the positivity of the operator
${\mathbf M}$ is the fact that the final states are less mutually
distinguishable than the initial states.  That is,  the modulus of
the overlap of each pair of final states is greater than that of
their initial counterparts.   From Eq. (5), we see that this is
equivalent to the condition
\begin{equation}
|{\mu}_{jj'}|{\le}1.
\end{equation}
To derive this from the positivity of ${\mathbf M}$, consider the
normalised vector ${\mathbf v}^{jj'}{\in}H_{N}$, where
$j{\neq}j'$, whose components are
\begin{equation}
v^{jj'}_{k}=\frac{1}{\sqrt
2}({\delta}_{jk}-e^{-i{\theta}_{jj'}}{\delta}_{j'k}),
\end{equation}
where ${\theta}_{jj'}={\mathrm Arg}({\mu}_{jj'})$. For this
vector, we find
\begin{equation}
{\langle}{\mathbf v}^{jj'},{\mathbf M}{\mathbf
v}^{jj'}{\rangle}=1-|{\mu}_{jj'}|.
\end{equation}
The positivity of ${\mathbf M}$ implies that this is non-negative,
leading to Eq. (12).  It is thus impossible for a deterministic
transformation of a set of pure states to result in any pair of
them becoming more distinguishable.

Quantum states cannot undergo deterministic transformations for
which ${\mathbf M}$ is not positive.  If, however, they are
linearly-independent, then they can undergo any transformation at
all with some non-zero probability. This follows from the fact
that linearly-independent states can be probabilistically
discriminated\cite{Me1}.  If the state is successfully identified,
then it can be transformed into any other state. General
probabilistic transformations of 2 pure states for which ${\mathbf
M}$ is not positive are examined fully in \cite{Me2}.

We have obtained a necessary, and for linearly-independent initial
states, sufficient condition for the existence of a CPLTP map
which deterministically transforms one set of pure quantum states
into another.  What about the sufficient conditions for
linearly-dependent states?  A more general problem is the
following: given a CPLTP map ${\cal L}$ which transforms one set
of linearly-independent pure states $|{\psi}^{1}_{j}{\rangle}$
into some other set of states $|{\psi}^{2}_{j}{\rangle}$, under
what circumstances is a pure state superposition of the initial
states, which we may call $|{\phi}^{1}{\rangle}$, transformed into
another pure state, $|{\phi}^{2}{\rangle}$?

When the $|{\psi}^{2}_{j}{\rangle}$ are also linearly-independent
this question is answered by

\begin{theorem}  Let $\{|{\psi}^{1}_{j}{\rangle}\}$ be a set
of linearly-independent quantum states spanning a complex Hilbert
space ${\cal H}$, and let there be a CPLTP map ${\cal L}$ which
transforms this set into some other set of linearly-independent
pure states $\{|{\psi}^{2}_{j}{\rangle}\}$. If there exists a pure
state
\begin{equation}
|{\phi}^{1}{\rangle}=\sum_{j}q_{j}|{\psi}^{1}_{j}{\rangle},
\end{equation}
in ${\cal H}$, where all $q_{j}{\neq}0$, such that ${\cal
L}(|{\phi}^{1}{\rangle}{\langle}{\phi}^{1}|)=|{\phi}^{2}{\rangle}{\langle}{\phi}^{2}|$,
for some pure state $|{\phi}^{2}{\rangle}$, then there exists a
unitary operator $U$ such that
\begin{equation}
|{\psi}^{2}_{j}{\rangle}{\langle}{\psi}^{2}_{j}|=U|{\psi}^{1}_{j}{\rangle}{\langle}{\psi}^{1}_{j}|U^{\dagger}.
\end{equation}
\end{theorem}
{\noindent}{\bf Proof:} We wish to examine the consequences of the
requirement that $|{\phi}^{1}{\rangle}$ is transformed into some
other pure state $|{\phi}^{2}{\rangle}$, which we may write as
\begin{equation}
|{\phi}^{2}{\rangle}=\sum_{j}r_{j}|{\psi}^{2}_{j}{\rangle},
\end{equation}
for some coefficients $r_{j}$.  It follows from Eqs. (3-6) that
\begin{equation}
\sum_{j,j'}q_{j'}^{*}q_{j}{\mu}_{jj'}|{\psi}^{2}_{j}{\rangle}{\langle}{\psi}^{2}_{j'}|=\sum_{j,j'}r_{j'}^{*}r_{j}|{\psi}^{2}_{j}{\rangle}{\langle}{\psi}^{2}_{j'}|.
\end{equation}
One consequence of the linear-independence of the
$|{\psi}^{2}_{j}{\rangle}$ is that
$q^{*}_{j'}q_{j}{\mu}_{jj'}=r^{*}_{j'}r_{j}$.  This can be seen
using an {\em orthogonalisation} operator $Q$, which acts
according to $Q|{\psi}^{2}_{j}{\rangle}=|x_{j}{\rangle}$, for some
orthonormal basis states $|x_{j}{\rangle}$.  Such an operator
always exists for linearly-independent states\cite{Me1}.  Acting
upon both sides of Eq. (18) with the superoperator
$Q(\;\;)Q^{\dagger}$, then taking matrix elements of both sides in
the $|x_{j}{\rangle}$ basis, gives this result, from which we
obtain
\begin{equation}
{\mu}_{jj'}=\left(\frac{r^{*}_{j'}}{q^{*}_{j'}}\right)\left(\frac{r_{j}}{q_{j}}\right).
\end{equation}
Thus, the operator ${\mathbf M}$ has unit rank.  Its Hermiticity and the fact that
${\mathrm Tr}({\mathbf M})=N$, which follows from Eq. (5), enable us to write ${\mathbf
M}=N{\mathbf P}$ for some projection operator ${\mathbf P}$.  It is convenient to write
\begin{equation}
{\mu}_{jj'}=Na^{*}_{j'}a_{j},
\end{equation}
where $a_{j}=r_{j}/N^{1/2}q_{j}$, and $\sum_{j}|a_{j}|^{2}=1$. It
follows from Eq. (5) that all of the diagonal elements of
${\mu}_{jj'}$ are equal to 1. Thus, the absolute values of the
$a_{j}$ must all be equal to $N^{-1/2}$. Consequently, we may
write $a_{j}=N^{-1/2}e^{i{\phi}_{j}}$, for some angles
${\phi}_{j}$. This gives
\begin{equation}
{\langle}{\psi}^{1}_{j'}|{\psi}^{1}_{j}{\rangle}=e^{i({\phi}_{j}-{\phi}_{j'})}{\langle}{\psi}^{2}_{j'}|{\psi}^{2}_{j}{\rangle},
\end{equation}
or equivalently,
\begin{equation}
|{\psi}^{2}_{j}{\rangle}=e^{-i{\phi}_{j}}U|{\psi}^{1}_{j}{\rangle},
\end{equation}
for some unitary operator $U$, leading to Eq. (16).  The overall
phases of the initial states are physically irrelevant, so we may
say that, physically, the initial and final states are related by
a unitary transformation.  This completes the proof.

It is a simple matter to generalise this theorem to the case when
only some (but at least 2) of the $q_{j}$ are non-zero.  If we
denote by $S$ the set of values of $j$ for which $q_{j}{\neq}0$,
then the result is that Eq. (16) holds for $j{\in}S$.  To see
this, consider only a deterministic transformation mapping the
$|{\psi}_{j}^{1}{\rangle}$, for $j{\in}S$, into the corresponding
final states $|{\psi}_{j}^{2}{\rangle}$.  Then all of the
assumptions of the theorem are satisfied, and the result follows.

Theorem 2 cannot hold if the condition of linear-independence of
the final states is dropped, for the simple reason that no unitary
transformation can transform a linear-independent set into a
linearly-dependent one.

To conclude, considerable efforts have recently been devoted to
understanding the conditions under which one set of states can be
{\em probabilistically} transformed into another.  This issue has
arisen, for example, in discussions of probabilistic
state-discrimination and cloning\cite{Me1,Me2,Duanguo}.  Yet
little has been known about the circumstances under which {\em
deterministic} state transformations are possible.  In this
Letter, we have solved the problem when the initial states are
linearly-independent.  We have also shown how, if both the initial
and final sets of states are linearly-independent, the constraint
of maintaining the purity of just one complete superposition means
that the transformation must be effectively unitary.  By
`effectively' unitary, we mean that the initial and final states
are related by a unitary transformation up to physically
irrelevant phases. Thus, while other, non-trivial deterministic
transformations between linearly-independent states may convert
one set of pure states into another, they will destroy the purity
of all other states.
\section*{Acknowledgements}
The author gratefully acknowledges the EPSRC for the award of a
research fellowship.  Thanks also go to Christof Zalka for his
useful comments.

\end{document}